\RequirePackage[undo-recent-deprecations]{expl3}
\documentclass[twoside,twocolumn,9pt]{article}
\usepackage{extsizes}
\usepackage[super,sort&compress,comma]{natbib} 
\usepackage[version=3]{mhchem}
\usepackage[left=1.5cm, right=1.5cm, top=1.785cm, bottom=2.0cm]{geometry}
\usepackage{balance}
\usepackage{mathptmx}
\usepackage{verbatim}
\usepackage{sectsty}
\usepackage{graphicx} 
\usepackage{lastpage}
\usepackage[format=plain,justification=justified,singlelinecheck=false,font={stretch=1.125,small,sf},labelfont=bf,labelsep=space]{caption}
\usepackage{float}
\usepackage{fancyhdr}
\usepackage{fnpos}
\usepackage[english]{babel}
\addto{\captionsenglish}{%
  
}
\usepackage{array}
\usepackage{droidsans}
\usepackage{charter}
\usepackage[T1]{fontenc}
\usepackage[usenames,dvipsnames]{xcolor}
\usepackage{setspace}
\usepackage[compact]{titlesec}
\usepackage{hyperref}

\usepackage{epstopdf}

\definecolor{cream}{RGB}{222,217,201}

\begin{document}

\pagestyle{fancy}
\thispagestyle{plain}
\fancypagestyle{plain}{
\renewcommand{\headrulewidth}{0pt}
}

\makeFNbottom
\makeatletter
\renewcommand\LARGE{\@setfontsize\LARGE{15pt}{17}}
\renewcommand\Large{\@setfontsize\Large{12pt}{14}}
\renewcommand\large{\@setfontsize\large{10pt}{12}}
\renewcommand\footnotesize{\@setfontsize\footnotesize{7pt}{10}}
\makeatother

\renewcommand{\thefootnote}{\fnsymbol{footnote}}
\renewcommand\footnoterule{\vspace*{1pt}%
\color{cream}\hrule width 3.5in height 0.4pt \color{black}\vspace*{5pt}} 
\setcounter{secnumdepth}{5}

\makeatletter 
\renewcommand\@biblabel[1]{#1}            
\renewcommand\@makefntext[1]%
{\noindent\makebox[0pt][r]{\@thefnmark\,}#1}
\makeatother 
\renewcommand{\figurename}{\small{Fig.}~}
\sectionfont{\sffamily\Large}
\subsectionfont{\normalsize}
\subsubsectionfont{\bf}
\setstretch{1.125} 
\setlength{\skip\footins}{0.8cm}
\setlength{\footnotesep}{0.25cm}
\setlength{\jot}{10pt}
\titlespacing*{\section}{0pt}{4pt}{4pt}
\titlespacing*{\subsection}{0pt}{15pt}{1pt}

\fancyfoot{}
\fancyfoot[LO,RE]{\vspace{-7.1pt}\includegraphics[height=9pt]{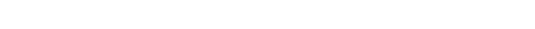}}
\fancyfoot[CO]{\vspace{-7.1pt}\hspace{11.9cm}\includegraphics{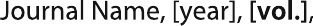}}
\fancyfoot[CE]{\vspace{-7.2pt}\hspace{-13.2cm}\includegraphics{head_foot/RF}}
\fancyfoot[RO]{\footnotesize{\sffamily{1--\pageref{LastPage} ~\textbar  \hspace{2pt}\thepage}}}
\fancyfoot[LE]{\footnotesize{\sffamily{\thepage~\textbar\hspace{4.65cm} 1--\pageref{LastPage}}}}
\fancyhead{}
\renewcommand{\headrulewidth}{0pt} 
\renewcommand{\footrulewidth}{0pt}
\setlength{\arrayrulewidth}{1pt}
\setlength{\columnsep}{6.5mm}
\setlength\bibsep{1pt}

\makeatletter 
\newlength{\figrulesep} 
\setlength{\figrulesep}{0.5\textfloatsep} 

\newcommand{\topfigrule}{\vspace*{-1pt}%
\noindent{\color{cream}\rule[-\figrulesep]{\columnwidth}{1.5pt}} }

\newcommand{\botfigrule}{\vspace*{-2pt}%
\noindent{\color{cream}\rule[\figrulesep]{\columnwidth}{1.5pt}} }

\newcommand{\dblfigrule}{\vspace*{-1pt}%
\noindent{\color{cream}\rule[-\figrulesep]{\textwidth}{1.5pt}} }

\makeatother

\twocolumn[
  \begin{@twocolumnfalse}
{\includegraphics[height=30pt]{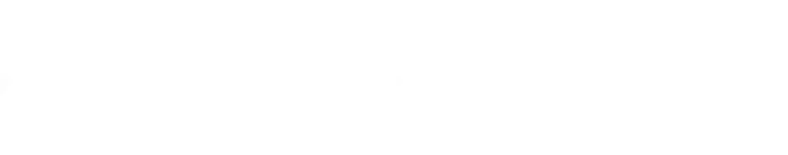}\hfill\raisebox{0pt}[0pt][0pt]{\includegraphics[height=55pt]{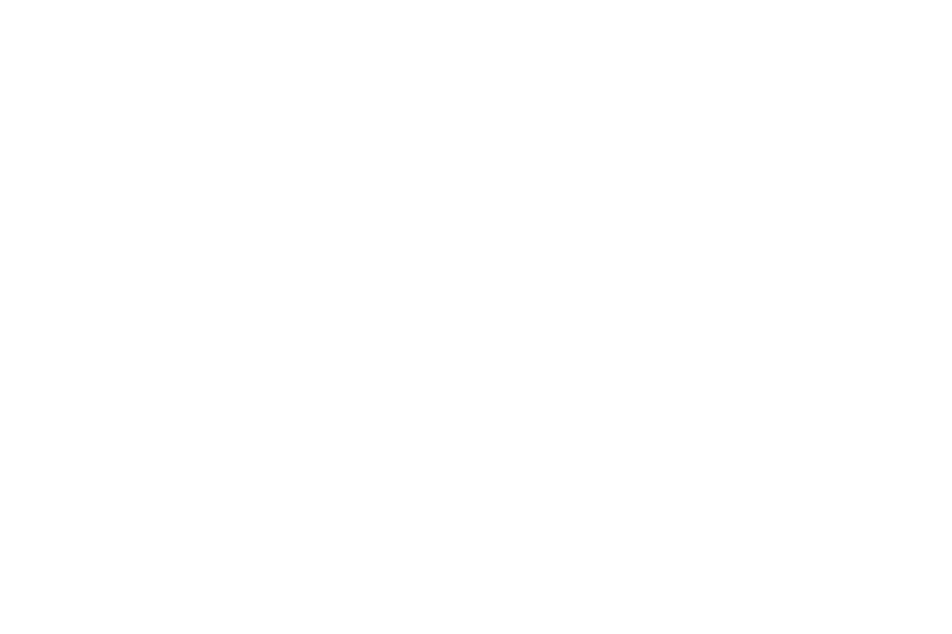}}\\[1ex]
\includegraphics[width=18.5cm]{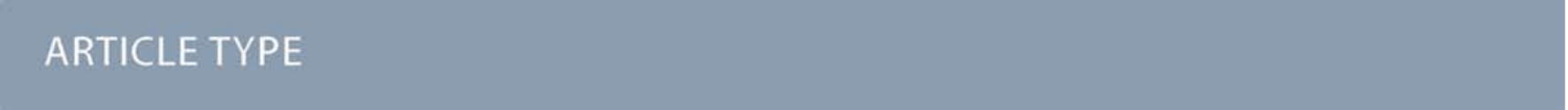}}\par
\vspace{1em}
\sffamily
\begin{tabular}{m{4.5cm} p{13.5cm} }

\includegraphics{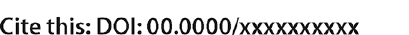} & \noindent\LARGE{\textbf{Self-Folding and Self-Scrolling Mechanisms of Edge-Deformed Graphene Sheets: A Molecular Dynamics Study}} \\
\vspace{0.3cm} & \vspace{0.3cm} \\

 & \noindent\large{Marcelo Lopes Pereira Junior$^{a}$, and Luiz Antonio Ribeiro Junior$^{a,*}$} \\

\includegraphics{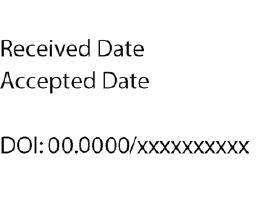} & \noindent\normalsize{Graphene-based nanofolds (GNFs) are edge-connected 2D stacked monolayers originated from single-layer graphene. Graphene-based nanoscrolls (GNSs) are nanomaterials with geometry resembling graphene layers rolled up into a spiral (papyrus-like) form. Both GNSs and GNFs structures induce significant changes in the mechanical and optoelectronic properties of single-layer graphene, aggregating new functionalities in carbon-based applications. Here, we carried our fully atomistic reactive (ReaxFF) molecular dynamics simulations to study the self-folding and self-scrolling of edge-deformed graphene sheets. We adopted initial armchair edge-scrolled graphene (AESG($\phi$,$\theta$)) structures with similar (or different) twist angles ($\phi$,$\theta$) in each edge, mimicking the initial configuration that was experimentally developed to form biscrolled sheets. Results showed that AESG(0,$2\pi$) and AESG(2$\pi$,$2\pi$) evolved to single-folded and two-folded fully stacked morphologies, respectively. As a general trend, for twist angles higher than $2\pi$, the self-deformation process of AESG morphologies yields GNSs. Edge twist angles lower than $\pi$ are not enough for triggering the self-deformation processes. In the AESG(0,3$\pi$) and AESG(3$\pi$,3$\pi$) cases, after a relaxation period, their morphology transition towards GNS occurred rapidly. In the AESG(3$\pi$,3$\pi$) dynamics, a metastable biscroll was formed by the interplay between the left- and right-sided partial scrolling in forming a unique GNS. At high-temperature perturbations, the edge folding and scrolling transitions to GNFs and GNSs occurred within the ultrafast period. Remarkably, the AESG(2$\pi$,3$\pi$) evolved to a dual state that combines folded and scrolled structures in a temperature-independent process.} \\

\end{tabular}

 \end{@twocolumnfalse} \vspace{0.6cm}

  ]

\renewcommand*\rmdefault{bch}\normalfont\upshape
\rmfamily
\section*{}
\vspace{-1cm}


\footnotetext{\textit{$^a$ Institute of Physics, University of Bras\'{i}lia, Bras\'{i}lia 70919-970, Brazil.}}
\footnotetext{Corresponding Author: ribeirojr@unb.com (L.A.R.J.)}
\footnotetext{\dag~Electronic Supplementary Information (ESI) available: The Supplementary Material presents the videos for MD simulations of all the systems studied here.}




\section{Introduction}

\begin{figure*}[!htb]
	\centering
	\includegraphics[width=0.8\linewidth]{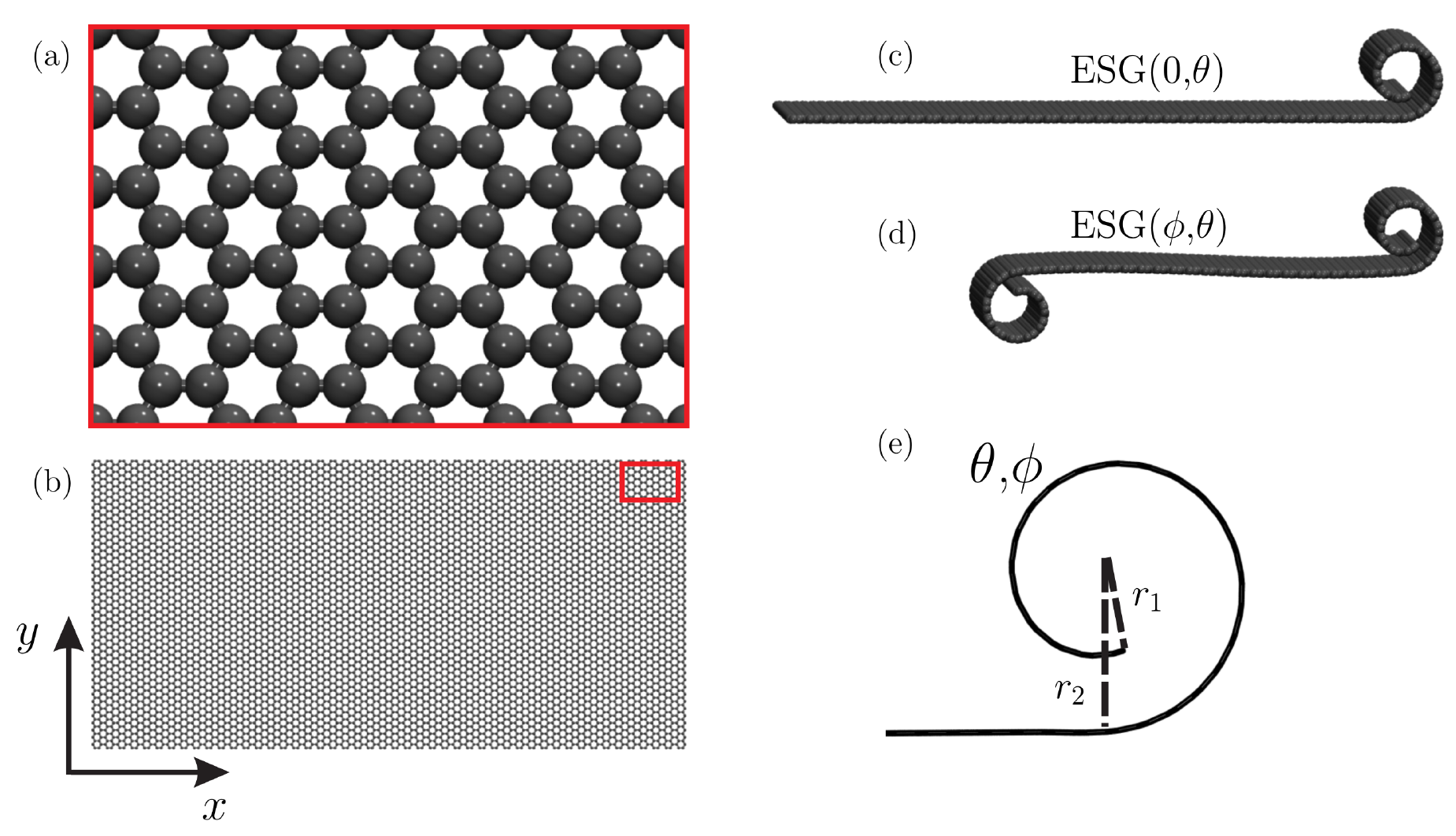}
	\caption{Schematic representation of the model AESG lattices studied here.(a) zooms-in a region of the adopted graphene sheet, which is presented in (b). (c) and (d) illustrate, respectively, the one-edge-scrolled graphene (AESG(0,$\theta$)) and two-edge-scrolled graphene (AESG($\phi$,$\theta$)) models used as input structures. (e) shows the edge structure configuration with its inner ($r_1$) and outer ($r_2$) radius. $\phi$ can assume the following values: $-3\pi$, $-2\pi$, 0, $2\pi$, and $3\pi$. $\theta$ can assume the following values: 0, $2\pi$, and $3\pi$.}
	\label{fig:systems}
\end{figure*}

Graphene has been evolved into a field of unprecedented activity in physics, chemistry, and materials research \cite{zhu2010graphene}. Several of its unique properties originate from its two-dimensional layer of one-atom thickness \cite{neto2009electronic}. Due to its excellent in-plane mechanical flexibility and strong inter-layer van der Waals interactions, single-layer graphene (SLG) is susceptible to deforming into scrolled \cite{viculis2003chemical} and folded \cite{zhang2010free} configurations. Graphene nanoscrolls (GNSs) are composed of tubularly scrolled SLG, yielding a 1D topological structure \cite{braga2004structure,toma1993stability,lavin2002scrolls,tomanek2002mesoscopic}. Some GNSs capabilities, such as superlubricity and large current sustainability, surpassing the properties of SLG itself \cite{li2018superior}. Graphene nanofolds (GNFs), in turn, are edge-connected 2D stacked monolayers that also originated from SLG. Their continuous and curved edge --- with a twist angle ($\theta$) between the two membranes --- can exhibit different band structures $\theta$-dependent interfacial conductivity \cite{wang2017controlled}. Both GNSs and GNFs structures induce significant changes in the mechanical and optoelectronic properties of SLG, aggregating new functionalities in graphene-based applications \cite{lima2011biscrolling,cranford2009meso,berman2015macroscale,patra2009nanodroplet,pereirajrPCCP_2020,tromer_PE}. 

Several experimental and theoretical studies have been performed to elucidate the self-scrolling \cite{berman2015macroscale,braga2004structure,shi2010mechanics,shi2010translational,xia2010fabrication,chu2011fabrication} and self-folding \cite{chen2014graphene,wang2017controlled,kim2011multiply,meng2013mechanics,mu2015origami,shenoy2012self} processes of SLG. Experimentally, high-resolution atomic force microscopy (AFM) \cite{chen2014graphene,chen2012novel}, scanning electron microscopy (SEM) \cite{yoo2012graphene,zheng2016preparation}, and transmission electron microscopy (TEM) \cite{zhang2010free,berman2015macroscale} have been employed in the study of SLG folding or scrolling deformations on flat substrates or in suspended conditions. In a theoretical perspective, continuum mechanics (CM) and molecular dynamics (MD) simulation techniques were employed to also investigate the SLG folding and scrolling behaviors \cite{braga2004structure,bellido2010molecular,zheng2011mechanical,gao2020molecular,zhu2013hydrogenation,cranford2009meso}. Importantly, these theoretical studies provide useful insights on the GNS and GNF formations, some of which may not be readily obtainable in experimental measurements (e.g., graphene local folding curvatures or the potential energy landscape for the non-bonded interaction terms). As a general trend, these works concluded that the self-formation of both GNSs and GNFs results from a interplay between the SLG bending rigidity and the van der Waals interactions. 

Different analytical models \cite{li2018mechanics,cox2015relating} and experimental setups \cite{lima2011biscrolling} have been proposed to study the scrolling and folding mechanics of SLG. Recently, Li \textit{et. al} proposed a finite-deformation analytical model in which no presumed assumptions on the geometries of deformed configurations are required \cite{li2018mechanics}. Both scroll-predicted and fold-predicted profiles for SLG and their critical conditions showed good agreement with MD simulation results. Lima \textit{et. al} developed an experimental approach that involves a twist-based spinning of carbon-based layers to form a biscrolled sheet \cite{lima2011biscrolling}. In that case, they converted a carbon nanotube sheet into a biscrolled nanotube sheet. The initial setup for the experiment was based on a structure that presents a cross-section similar to the illustration in Figure \ref{fig:systems}(d). Their findings revealed that the ability to make biscrolled yarns can be exploited to optimize yarn properties. Despite the amount of successful studies on the SLG conversion into GNSs and GNFs, their formation from a edge-deformed SLG is not fully understood.   

The self-folding \cite{shinNANOSCALE_2014} and self-scrolling \cite{meyerNATURE_2007} mechanisms of edge-deformed graphene sheets were experimentally reported recently. Shin and coworkers \cite{shinNANOSCALE_2014} introduced an ice-templated self-assembly approach for integrating two-dimensional graphene nanosheets to fabricate graphene nanoscroll networks. They showed that freeze-casting of reduced graphene oxide solution results in the formation of graphene nanoscroll networks at pH 10. In addition, they have demonstrated that graphene nanoscroll networks show good electrocatalytic activity for the oxygen reduction reaction. In the synthesis of graphene nanoscrolls, the self-scrolling process started from an edge-curved graphene sheet that triggers the self-scrolling process of graphene, producing nanoscrolls stabilized by the $\pi$-$\pi$ interactions. Meyer \textit{et al.} \cite{meyerNATURE_2007} observed folded edges in the fabrication of monolayer and bilayer graphene. During their fabrication processes, one of the edges was curved, forming graphene folds spontaneously. Importantly, these investigations do not provide information on the time needed for accomplishing the self-scrolling or self-folding of edge-deformed graphene and for the possible products yielded in these processes. In this sense, our investigation can shed light on the self-deformation of curved graphene sheets contributing to their better understanding.

In the present work, motivated by the deformation mechanism of carbon-based sheets proposed by Lima \textit{et al} \cite{lima2011biscrolling}, the self-folding and self-scrolling of a edge-deformed SLG was investigated in the framework of fully-atomistic reactive molecular dynamics simulations. Our computational approach was based on adopting initial armchair edge-scrolled graphene (AESG) structures with similar (or different) twist angles in each edge, mimicking the initial configuration used to form biscrolled sheets reported in reference \cite{lima2011biscrolling}. Moreover, we also provided information on the folding and scrolling curvatures and the potential energy landscapes during the AESG deformation processes.   

\section{Methodology}

To study the self-folding and self-scrolling of AESG sheets, we carried out fully-atomistic MD simulations within the scope of the reactive force field ReaxFF potential \cite{vanduin_JPCA,mueller_JPCC}, as implemented in the Large-scale Atomic/Molecular Massively Parallel Simulator (LAMMPS) \cite{plimpton_JCP}. Specifically, we used the parameters set provided by Chenoweth \textit{et al.} for C/H/O \cite{chenoweth2008reaxff}.  

The initial configurations of our model AESG systems are depicted in Figure \ref{fig:systems}. Figure \ref{fig:systems}(a) zooms-in a region of the graphene sheet used here, which is presented in Figure \ref{fig:systems}(b). Figures \ref{fig:systems}(c) and \ref{fig:systems}(d) illustrate, respectively, the one-edge-scrolled graphene (AESG(0,$\theta$)) and two-edge-scrolled graphene (AESG($\phi$,$\theta$)) models used in our simulations as input structures. $\theta$ and $\phi$ allow different twist angles on each edge. Figure \ref{fig:systems}(e) shows the edge structure configuration with its inner ($r_1$) and outer ($r_2$) radius. The graphene sheet has $x$ and $y$ dimensions of 100 \r{A} and 200 \r{A}, respectively, totaling 7776 atoms. These structures were atomistically modeled using the Sculptor plug-in \cite{humphrey_JMG}, as implemented in the software Visual Molecular Dynamics (VMD) \cite{humphrey_JMG}. VMD was also used here to obtain the MD snapshots and trajectories. 

The equations of motion were numerically integrated using the velocity-Verlet integrator with a time-step of $0.25$ fs until 250 ps. We adopted an NVT ensemble and Nos\'e-Hoover thermostat \cite{hoover1985canonical}, with a time constant of 100 fs, to analyze the influence of temperature on the scrolling and folding behaviors. The temperature regimes ranged from 100 K up to 1000 K with an increment of 100 K. Initial atomic velocities were randomly generated for each temperature.   

\begin{figure*}[htb!]
	\centering
	\includegraphics[width=1.0\linewidth]{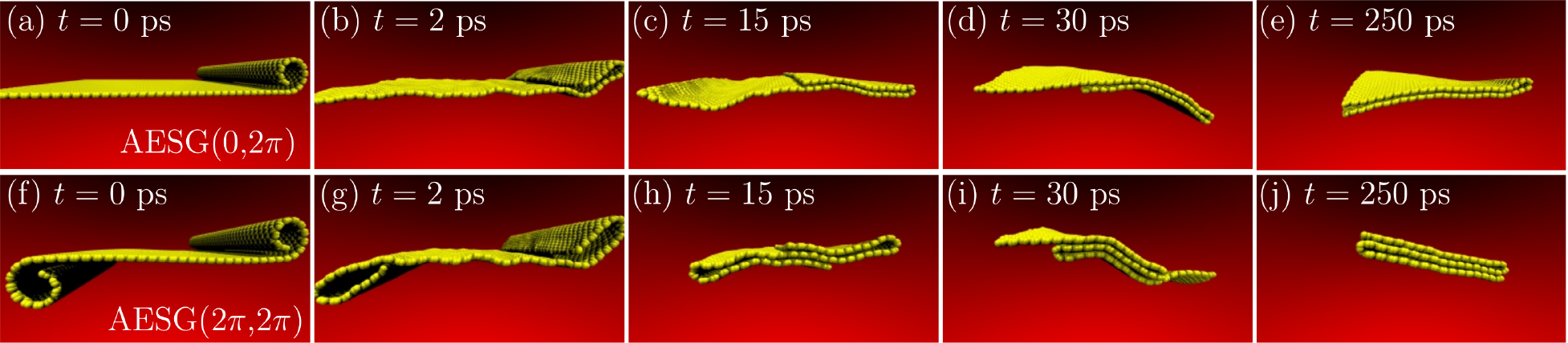}
	\caption{Representative MD snapshots for the self-deformation process of (a-e) AESG(0,2$\pi$) and (f-j) AESG(2$\pi$,2$\pi$) at 300 K. Importantly, the panels in this figure illustrate the slide motion obtained during the fold formation. In panels (c) and (d), the graphene edge on top moves towards the left direction, whereas the graphene edge at the bottom moves towards the right one. This trend of motion is better understood by watching the supplementary video SM1.}
	\label{fig:AESG-2pi}
\end{figure*}

\begin{figure*}[htb!]
	\centering
	\includegraphics[width=1.0\linewidth]{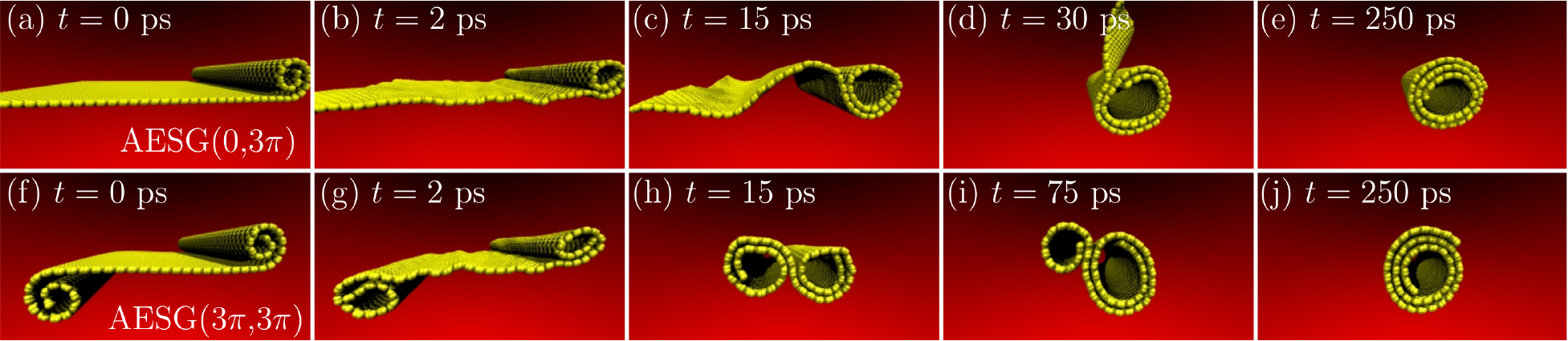}
	\caption{Representative MD snapshots for the self-deformation process of (a-e) AESG(0,3$\pi$) and (f-j) AESG(3$\pi$,3$\pi$) at 300 K.}
	\label{fig:AESG-3pi}
\end{figure*}

\section{Results}

\begin{figure*}[htb!]
	\centering
	\includegraphics[width=\linewidth]{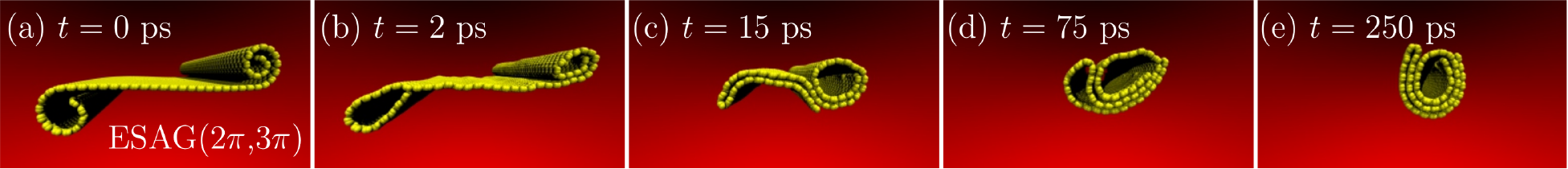}
	\caption{Representative MD snapshots for the self-deformation process of AESG($2\pi$,3$\pi$) at 300 K.}
	\label{fig:AESG-2pi-3pi}
\end{figure*}

We begin our discussion by presenting the MD snapshots for the folding and scrolling mechanisms of the model AESG systems. The Supplementary Material presents the videos for MD simulations of all the systems studied here at 300 K and 1000 K. Figures \ref{fig:AESG-2pi}(a-e) (and supplementary video SM1) and \ref{fig:AESG-2pi}(f-j) (and supplementary video SM2) illustrate the representative MD snapshots for the self-deformation process of AESG(0,2$\pi$) and AESG(2$\pi$,2$\pi$) cases at 300 K, respectively. In Figures \ref{fig:AESG-2pi}(a) and \ref{fig:AESG-2pi}(b) one can note that a smooth morphology transition, from a twisted edge towards a folded one, occurs within an ultrafast time-period, about 2 ps. It is the period necessary for the system to escape from the local potential well at this specific thermal excitation. For higher temperature perturbations, the edge folding and scrolling transitions occur much faster. At 1000 K, the transition mechanism illustrated in Figures \ref{fig:AESG-2pi}(a-e) also takes place, but before 1 ps (see supplementary video SM3). During 240 ps, the staked monolayers formed in the morphology transition slide against each other, as shown in Figures \ref{fig:AESG-2pi}(c) and \ref{fig:AESG-2pi}(d). After this transient period, a fully GNF is formed at 250 ps (Figure \ref{fig:AESG-2pi}(e)). It is worth noting that this fully single-folded structure was also obtained at 1000 K (see supplementary video SM3). For the AESG(2$\pi$,2$\pi$) case (Figures \ref{fig:AESG-2pi}(f-j)), the edge scrolling-folding transition occurs in the same fashion, but a two-folded structure was yielded instead. This two-folded structure is only formed below a critical temperature regime, about 900 K. Above this temperature threshold, the random motion of the atoms, imposed by the thermal fluctuations, overcomes the potential energy well (bond energies plus van der Waals interactions) related to the formation of the two-folded structure, and it cannot overcome the potential energy well of a single-folded one. In this way, the edge scrolling-folding transition at 1000 K leads to a fully single-folded morphology (see supplementary video SM4).  

Figures \ref{fig:AESG-3pi}(a-e) (and supplementary video SM5) and \ref{fig:AESG-3pi}(f-j) (and supplementary video SM6) illustrate the representative MD snapshots for the self-deformation process of AESG(0,3$\pi$) and AESG(3$\pi$,3$\pi$) cases at 300 K, respectively. As a general trend, for twist angles higher than $2\pi$, the self-deformation process of AESG yields GNSs morphologies. Edge twist angles lower than $\pi$ are not enough for triggering the self-deformation processes once the interlayer van der Waals interactions in the edge can not overcome the SLG bending rigidity. Consequently, the AESG structure is converted into a conventional SLG. As shown in Figures \ref{fig:AESG-3pi}(a) and \ref{fig:AESG-3pi}(b), the AESG(0,3$\pi$) leads about 2 ps to relax into a form that roughly preserves the initial size of the scrolled part. After this relaxation period, the transition from AESG(0,3$\pi$) towards a GNS morphology occurs rapidly (about 30 ps), according to Figures \ref{fig:AESG-3pi}(c) and \ref{fig:AESG-3pi}(d). At about 50 ps, a stable GNS with entirely overlapped layers is formed and keeps its integrity until the end of the simulation (Figure \ref{fig:AESG-3pi}(e)). As expected, Figures \ref{fig:AESG-3pi}(f) and \ref{fig:AESG-3pi}(g) show that the self-deformation process of AESG(3$\pi$,3$\pi$) presents a relaxation period similar to the AESG(3$\pi$,3$\pi$) case. At 15 ps, a metastable biscroll is formed by the interplay between the left- and right-sided partial scrolling in forming a unique GNS, as illustrated in Figure \ref{fig:AESG-3pi}(h). At 75 ps, the right side wins and a single GNS starts to be formed (see Figure \ref{fig:AESG-3pi}(i)) and a GNS with entirely overlapped layers is formed at 76 ps, remaining stable until the simulation, as depicted in Figure \ref{fig:AESG-3pi}(j).

Note that there is no preferred side in converting a metastable biscroll into a GNS. This is a random process that depends on the seed used to randomly generate the atom velocities in LAMMPS. Moreover, the thermal fluctuations imposed on the displacement of the atoms are randomly obtained and can also impact the conversion process of a metastable biscroll into a GNS. Differently to what was obtained for the folding cases (Figure \ref{fig:AESG-2pi}), the AESG(0,3$\pi$) and AESG(3$\pi$,3$\pi$) MD formed stable GNSs even at high temperatures. See supplementary videos SM7 and SM8 for the AESG(0,3$\pi$) and AESG(3$\pi$,3$\pi$) MD at 1000 K, respectively. In the supplementary videos SM5-SM8, one can note that the GNS presents a breathing motion that varies the inner radius amplitude in an oscillatory fashion. This behavior was detailed discussed in references \cite{shi2009gigahertz}.

We now present in Figure \ref{fig:AESG-2pi-3pi} (and in the supplementary video SM9) the representative MD snapshots for the AESG(2$\pi$,3$\pi$) case, in which the edges were initially scrolled with different twist angles. Remarkably, the AESG(2$\pi$,3$\pi$) evolves to a dual state that combines folded and scrolled structures. This process is temperature-independent (see supplementary videos SM9 and SM10 (AESG(2$\pi$,3$\pi$) at 1000 K)). The AESG(2$\pi$,3$\pi$) relaxation process illustrated in Figures \ref{fig:AESG-2pi-3pi}(a) and \ref{fig:AESG-2pi-3pi}(b) occurs according to the mechanisms described above for the AESG(0,2$\pi$) and AESG(0,3$\pi$) cases (Figures \ref{fig:AESG-2pi}) and \ref{fig:AESG-3pi}, respectively). In Figure \ref{fig:AESG-2pi-3pi}(c), one can realize that edge initially scrolled with $2\pi$ tends to form a GNF whereas the edge scrolled with $3\pi$ tries to form a GNS. At 75 ps, the partially formed fold and scroll overlap preserving their forms (see Figure \ref{fig:AESG-2pi-3pi}(d)). The final structure obtained in the AESG(2$\pi$,3$\pi$) MD is presented in Figure \ref{fig:AESG-2pi-3pi}(d). It is worth mentioning that the coexistence of folded and scrolled shapes of graphene in the final morphology presented Figure \ref{fig:AESG-2pi-3pi}(d) has not been reported so far.         

\begin{figure}[htb!]
	\centering
	\includegraphics[width=0.9\linewidth]{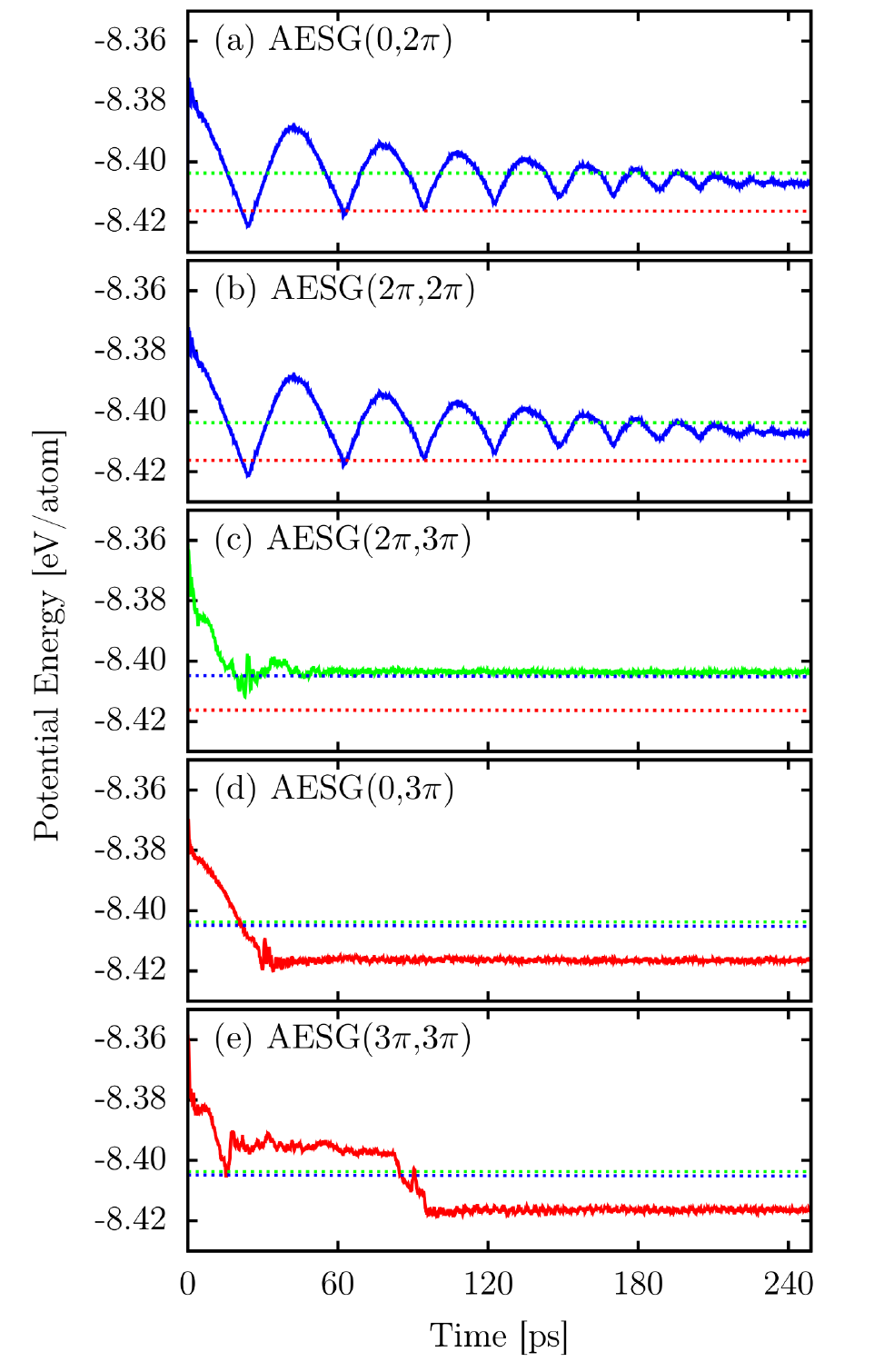}
	\caption{Time evolution of the potential energy for MD simulations presented above: (a) AESG(0,2$\pi$), (b) AESG(2$\pi$,2$\pi$), (c) AESG(2$\pi$,3$\pi$), (d) AESG(0,3$\pi$), and (e) AESG(3$\pi$,3$\pi$).}
	\label{fig:energies}
\end{figure}

In the context of our simulations, the edge topology does not affect the folding and scrolling mechanisms of the graphene sheets. The Supplementary Material videos SM11-SM15 show the dynamics of zigzag ESG (ZESG) at 300 K, considering all the initial structures modeled here. The supplementary videos SM11, SM12, SM13, SM14, and SM15 are referring to the ZESG(0,2$\pi$), ZESG(2$\pi$,2$\pi$), ZESG(2$\pi$,3$\pi$), ZESG(0,3$\pi$), and ZESG(3$\pi$,3$\pi$) cases, respectively. By contrasting the results presented in the supplementary videos for the armchair cases at 300 K (SM1, SM3, SM5, SM7, and SM9) with the ones for the zigzag lattices (SM11-SM15), one can note that the zigzag and armchair ESG dynamics are similar (i.e., not edge-sensitive).

We carried out simulations considering negative values of $\phi$. These simulations are presented in the Supplementary Material (videos SM16-SM18) alongside their related plots for the time evolution of the potential energy. The supplementary videos SM16, SM17, and SM18 are referring to the AESG(-2$\pi$,2$\pi$), AESG(-3$\pi$,2$\pi$), and  AESG(-3$\pi$,3$\pi$) cases,respectively. For the AESG(-2$\pi$,2$\pi$) case, the system evolves to a collapsed structure with two-folded edges due to the formation of carbon-carbon covalent bonds since the edges are very close to each other. The AESG(-2$pi$,3$\pi$) case evolves to a single CNS, and the AESG(-3$\pi$,3$\pi$) case yields a dual state that combines folded and scrolled structures, similar to the results presented in Figure \ref{fig:AESG-2pi-3pi}(e). Figure S3 in Supplementary Material illustrates the representative MD snapshots for the AESG(-2$\pi$,2$\pi$) case, to highlight the collapsed structure with two-folded edges.

Finally, Figure \ref{fig:energies} shows the time evolution of the potential energy for all the cases discussed above. In this figure, the structural stability of formed folded, scrolled, and dual-species is characterized by the convergence of the curves. For the sake of comparison, the blue, red, and green dashed lines denote the value for the average potential energies (in the last 100 ps) for the AESG(0,2$\pi$), AESG(0,3$\pi$), and  AESG(2$\pi$,3$\pi$), respectively. Note that this average value is similar among the related AESG(0,$\theta$) and AESG($\theta$,$\theta$) cases. As expected, the curve profiles in Figure \ref{fig:energies} suggest that scrolled and folded configurations are energetically favorable. As observed in Figures \ref{fig:energies}(a) and \ref{fig:energies}(b), the potential energy oscillates in time due to the relative slide motion between the folded layers. This motion increases (or decreases) the van der Waals interactions depending on the stacked area in the GNF formation. The potential energy related to the folding processes is slightly higher than the stabilization energy in the scrolling processes (red dashed line) and slightly lower than the one obtained in forming the dual-species (green dashed line). In Figure \ref{fig:energies}(c), one can see that the potential energy in forming the dual state that combines folded and scrolled structures converges rapidly. Since the converged potential energy is marginally higher than the one for AESG(0,$\theta$) and AESG($\theta$,$\theta$) cases, the dual state is also energetically favorable. Figures \ref{fig:energies}(d) and \ref{fig:energies}(e) show that, during the dynamical process for the GNSs formation, there is a downhill trend for the potential energy values that ends about 40 ps and 100 ps for AESG(0,3$\theta$) and AESG(3$\theta$,3$\theta$), respectively. From that moment on, the total potential energy stabilizes and a flat region persists until the end of the simulation. As mentioned above, the collapsed CNSs are formed within the ultrafast time scale. The transient plateau (20-100 ps) in Figure \ref{fig:energies}(e) refers to the formation of the metastable biscroll (see Figure \ref{fig:AESG-3pi}(h)). At high temperatures, GNFs and GNSs formation energies are greater when compared, for instance, with results in Figure \ref{fig:energies}. 

\section{Conclusions}

In summary, we carried out fully atomistic (ReaxFF) MD simulations to study the self-folding and self-scrolling of a edge-deformed graphene sheet. We adopted initial edge-scrolled graphene structures with similar (or different) twist angles in each edge, mimicking the initial configuration used to form biscrolled sheets reported in reference \cite{lima2011biscrolling}. Results showed that AESG(0,$2\pi$) and AESG(2$\pi$,$2\pi$) evolved to single-folded and two-folded fully stacked morphologies, respectively. The two-folded structure is only formed below a critical temperature regime, about 900 K. The edge scrolling-folding transition of AESG(2$\pi$,$2\pi$), at 1000 K, leads to a fully single-folded morphology. 

As a general trend, for twist angles higher than $2\pi$, the self-deformation process of AESG morphologies yields GNSs. Edge twist angles lower than $\pi$ are not enough for triggering the self-deformation processes once the interlayer van der Waals interactions in the edge can not overcome the SLG bending rigidity. Consequently, the AESG structure is converted into a conventional graphene monolayer. In the AESG(0,3$\pi$) and AESG(3$\pi$,3$\pi$) cases, after a relaxation period, their morphology transitions towards GNS occurred rapidly. In the AESG(3$\pi$,3$\pi$) MD, a metastable biscroll was formed by the interplay between the left- and right-sided partial scrolling in forming a unique GNS. Differently to what was obtained for the folding cases (Figure \ref{fig:AESG-2pi}), the AESG(0,3$\pi$) and AESG(3$\pi$,3$\pi$) MD formed stable GNSs even at high temperatures. At high temperature perturbations, the edge folding and scrolling transitions to GNFs and GNSs occurred within an ultrafast period. Remarkably, the AAESG(2$\pi$,3$\pi$) evolves to a dual state that combines folded and scrolled structures. This process is temperature-independent. It is worth mentioning that the coexistence of folded and scrolled shapes of graphene in the final morphology presented Figure \ref{fig:AESG-2pi-3pi}(d) has not been reported so far.  

It is worth mentioning that some theoretical works in the literature carried out MD simulations using carbon nanotubes \cite{zhang2011ultrafast,huang2015mechanical,zhang2010carbon,perim2013controlled}, graphene nanoribbons \cite{wang2015formation}, diamond clusters \cite{berman2015macroscale}, and other kinds of agents \cite{bejagam2018nanoparticle,xia2010fabrication} for triggering the folding or scrolling processes of graphene. These works have not investigated the folding and scrolling mechanism triggered by edge deformations. Relevant experimental studies in literature reported self-folding and self-scrolling of graphene originated by its edge deformations \cite{shinNANOSCALE_2014,meyerNATURE_2007}. In this sense, our results shed light on this crucial mechanism. To the best of our knowledge, the conversion mechanism of edge-deformed graphene into folds and scrolls and the possible products yielded from this conversion --- such as the collapsed structure with two-folded edges and the dual state that combines folded and scrolled structures --- are first reported here.

\section*{Conflicts of interest}

There are no conflicts to declare.

\section*{Acknowledgements}
The authors gratefully acknowledge the financial support from Brazilian research agencies CAPES, CNPq, and FAP-DF. M.L.P.J. gratefully acknowledges the financial support from CAPES grant 88882.383674/2019-01. L.A.R.J. gratefully acknowledges the financial support from CNPq grant 302236/2018-0, FAP-DF grant $0193.0000248/2019-32$, DPI/DIRPE/UnB (Edital DPI/DPG $03/2020$) grant $23106.057541/2020-89$, and IFD/UnB (Edital $01/2020$) grant $23106.090790/2020-86$. The authors gratefully acknowledge the National Laboratory for Scientific Computing (LNCC/MCTI, Brazil) for providing HPC resources of the SDumont supercomputer, which have contributed to the research results reported within this work (http://sdumont.lncc.br). This research was developed with the support of CENAPAD-SP (Centro Nacional de Processamento de Alto Desempenho em S\~ao Paulo), grant UNICAMP / FINEP - MCT.



\balance


\bibliography{rsc} 
\bibliographystyle{rsc} 

\end{document}


\title{Supplementary Information for\\Self-Folding and Self-Scrolling Mechanisms of Edge-Deformed\\Graphene Sheets: A Molecular Dynamics Study}

\author{Marcelo Lopes Pereira Junior}
 \affiliation{Institute of Physics, University of Bras\'{i}lia, 70.919-970, Bras\'{i}lia, Brazil.}
 \author{Luiz Antonio Ribeiro Junior}
 \affiliation{Institute of Physics, University of Bras\'{i}lia, 70.919-970, Bras\'{i}lia, Brazil.}
\maketitle

\begin{figure}[!h]
\centering
\includegraphics[width=0.5\linewidth]{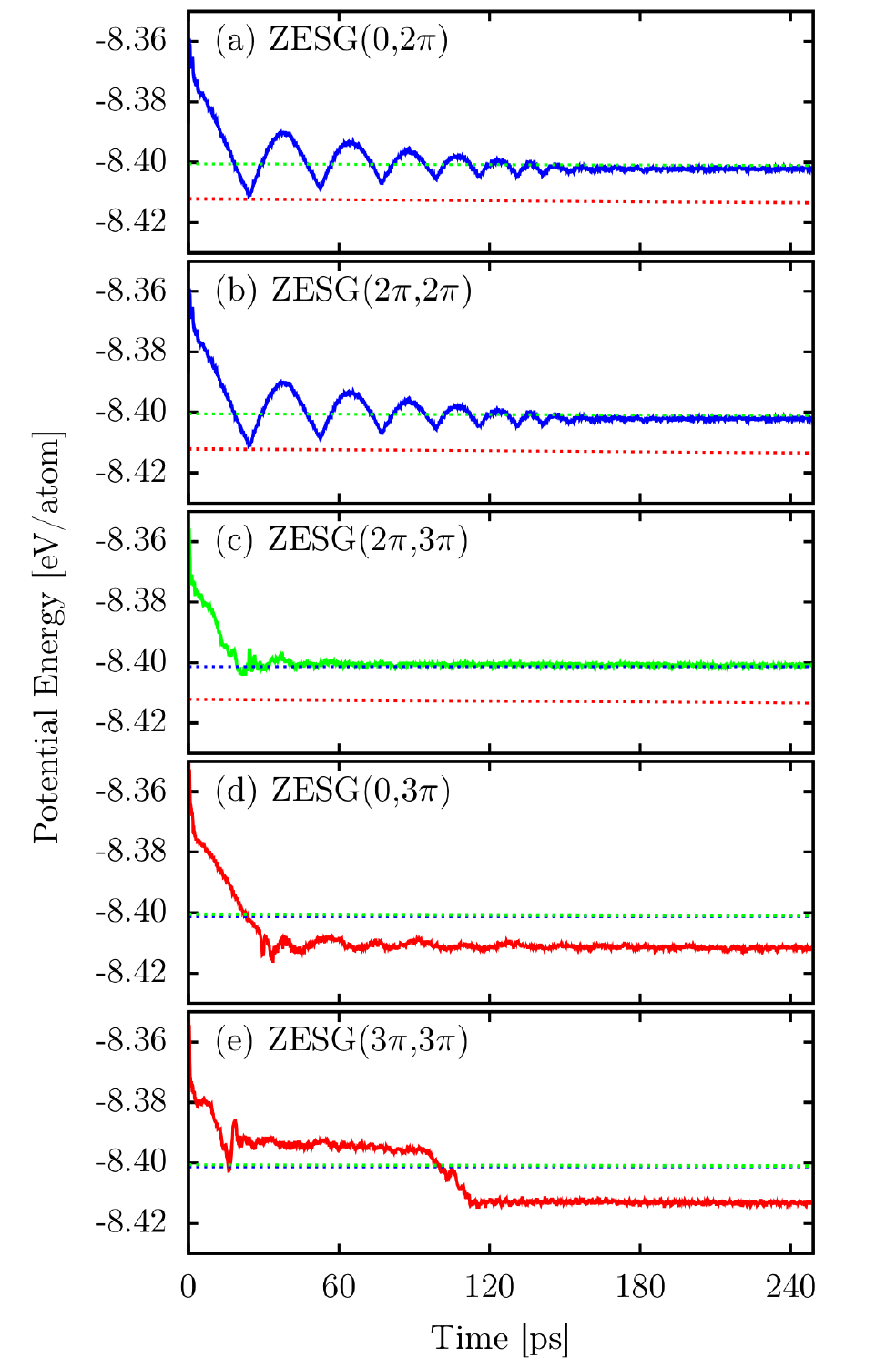}
\caption{Time evolution of the potential energy for the MD simulations of the following zigzag ESG sheets at 300 K: (a) ZESG(0,2$\pi$), (b) ZESG(2$\pi$,2$\pi$), (c) ZESG(2$\pi$,3$\pi$), (d) ZESG(0,3$\pi$), and (e) ZESG(3$\pi$,3$\pi$). For the sake of comparison, the blue, red, and green dashed lines denote the value for the average potential energies (in the last 100 ps) for the ZESG(0,2$\pi$), ZESG(0,3$\pi$), and ZESG(2$\pi$,3$\pi$), respectively. Note that this average value is similar among the related ZESG(0,$\theta$) $\theta$) cases. As expected, the curve profiles suggest that scrolled and folded configurations are energetically favorable also for the zigzag cases. Moreover, from this figure, it is also possible to conclude that the edge topology does not affect the folding and scrolling mechanisms of the graphene sheets since the trends presented by the time evolution of the potential energy are similar between the zigzag and armchair ESG cases.}
\label{figs1}
\end{figure}

\begin{figure}[!h]
\centering
\includegraphics[width=\linewidth]{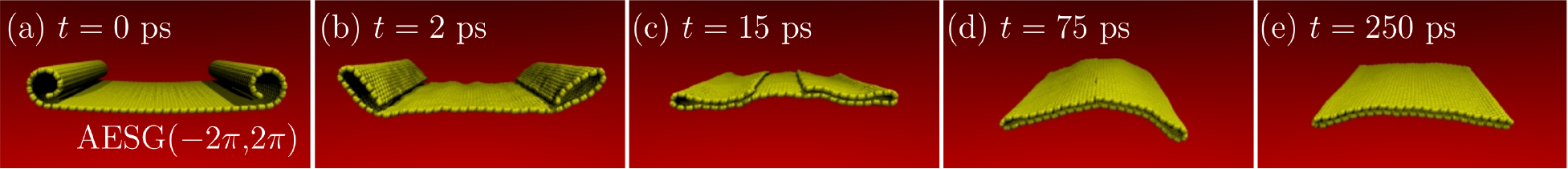}
\caption{Representative MD snapshots for the self-deformation process of AESG($-2\pi$,$2\pi$) at 300 K. The system evolves to a collapsed structure with two-folded edges due to the formation of carbon-carbon covalent bonds since the edges are very close to each other.}
\label{figs2}
\end{figure}

\begin{figure}[!h]
\centering
\includegraphics[width=0.5\linewidth]{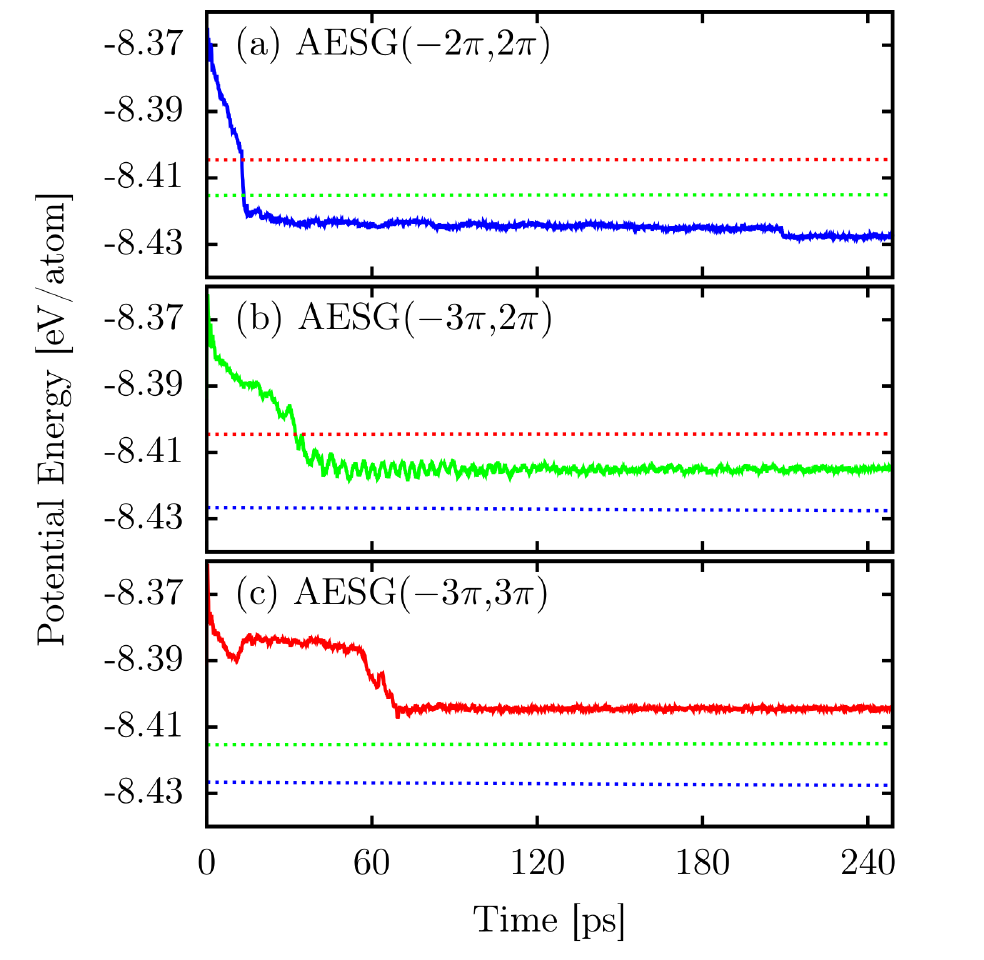}
\caption{Time evolution of the potential energy for the MD simulations of the following zigzag ESG sheets at 300 K: (a) AESG(-2$\pi$,2$\pi$), (b) AESG(-3$\pi$,2$\pi$), and (c) AESG(-3$\pi$,3$\pi$). For the sake of comparison, the blue, red, and green dashed lines denote the value for the average potential energies (in the last 100 ps) for the AESG(-2$\pi$,2$\pi$), AESG(-3$\pi$,2$\pi$), and AESG(-3$\pi$,3$\pi$), respectively. Note that this average value is similar among the related AESG(-$\phi$,$\theta$) cases. During the dynamical process of these structures, there is a downhill trend for the potential energy values that ends about 10 ps, 40 ps, and 60 ps for the AESG(-2$\pi$,2$\pi$), AESG(-3$\pi$,2$\pi$), and AESG(-3$\pi$,3$\pi$) cases, respectively. After these periods, the total potential energy stabilizes, and a flat region persists until the end of the simulation. The plateaus denote the formation of the final structures.}
\label{figs3}
\end{figure}